\documentclass[12pt, twoside]{article}
\usepackage{a4wide,amssymb,cite}
\usepackage{epsfig}

\newcommand{\be}{\begin{equation}}
\newcommand{\ee}{\end{equation}}
\newcommand{\bea}{\begin{eqnarray}}
\newcommand{\eea}{\end{eqnarray}}

\newcommand{\vep}{\varepsilon}

\def\a{\alpha}

\def\g{\gamma}
\def\G{\Gamma}
\def\d{\delta}

\def\la{\lambda}

\def\m{\mu}
\def\n{\nu}

\def\s{\sigma}

\def\f{\phi}

\def\vf{\varphi}
\def\ep{\epsilon}
\def\vep{\varepsilon}

\def\th{\theta}

\newcommand{\eg}{{\it e.g.,}\ }\newcommand{\ie}{{\it i.e.,}\ }
\newcommand{\reef}[1]{(\ref{#1})}

\begin{document}

\begin{titlepage}

\rightline{October 2007}

\begin{centering}
\vspace{2cm}
{\Large {\bf Supersymmetric codimension-two branes in six-dimensional gauged supergravity}}\\

\vspace{2cm}

 {\bf Hyun Min Lee}$^{a,*}$ and {\bf Antonios~Papazoglou}$^{b,**}$ \\
\vspace{.2in}

{\it $^{a}$ Department of Physics, Carnegie Mellon University, \\
5000 Forbes Avenue, Pittsburgh, PA 15213, USA.\\
\vspace{3mm}
$^{b}$ Institute of Cosmology \& Gravitation, University of Portsmouth,\\
Portsmouth PO1 2EG, UK.}\\

\end{centering}
\vspace{3cm}

\begin{abstract}
We consider the six-dimensional Salam-Sezgin supergravity in the
presence of codimension-2 branes. In the case that the branes
carry only tension, we provide a way to supersymmetrise them by
adding appropriate localised Fayet-Iliopoulos terms and localised corrections 
to the Chern-Simons term and modifying
accordingly the fermionic supersymmetry transformations. The resulting brane
action has ${\cal N}=1$ supersymmetry (SUSY). We find the
axisymmetric vacua of the system and show that one has unwarped
background solutions with "football"-shaped extra dimensions which
always respect ${\cal N}=1$ SUSY for any value of the equal brane tensions, 
in contrast with the non-supersymmetric brane action background. Finally, we generically
find multiple zero modes of the gravitino in this background and
discuss how one could obtain a single chiral zero mode present in
the low energy spectrum.
\end{abstract}

\vspace{2cm}

\begin{flushleft}

$^{*}~$ e-mail address: hmlee@andrew.cmu.edu \\
$ ^{**}$ e-mail address: antonios.papazoglou@port.ac.uk

\end{flushleft}

\end{titlepage}

\section{Introduction}

For the last decade, it has been an intensive effort to
incorporate gravity for solving the particle physics problems.
Particularly, in higher dimensional models with branes where the
Standard Model (SM) particles are confined \cite{branes}, the mass
scale hierarchies in the SM can be understood from geometric
factors in extra dimensions. Moreover, for the minimal
supersymmetric extension of the SM (MSSM), the SUSY flavor problem
can be ameliorated by a geometrical separation of the hidden
sector from the visible sector in extra dimensions, the so called
sequestering mechanism \cite{RS,LS}. In this case, the anomaly
mediation \cite{RS,anomalymed} can be a dominant
contribution\footnote{The K\"ahler potential is not of a
sequestered form in higher than five dimensions\cite{highseq,fll}
but some global symmetry that is not broken by the messenger
sector can keep the sequestering\cite{symseq}.} to the soft mass
parameters in the MSSM. The supersymmetric embedding of the brane
action in the 5D warped supergravity was studied in \cite{5dwarp}
and the extension of the analysis to the 6D flat supergravity has
been done in \cite{fll}.

Recently there has been a renewed interest into the 6D
Salam-Sezgin supergravity\cite{SS}, due to the findings of the new
warped solutions\cite{gibbons,otherwarped1,otherwarped,leelud,gian}. The warped
background has the extra dimensions ``spontaneously'' compactified
by $U(1)_R$ flux on the warped product of the 4D Minkowski space
and a deformed sphere (or general two-dimensional compact Riemann
surfaces). Moreover, the branes with nonzero tensions are
accommodated at the conical singularities, without the need of
cutting and pasting the extra dimension as in the 5D case. Since
the 4D Minkowski space is present as a unique regular solution
with maximal symmetry \cite{gibbons}, the warped solution has a
feature of self-tuning of the cosmological constant
\cite{6dself}\footnote{See, however, Refs.~\cite{finetune,finetune1}.} (for
a review, see \cite{koy}). There have been a lot of follow-up
works on this model (as well as its non-SUSY analogue
\cite{japs}), such as the perturbation analysis \cite{pert,salvio,
stabwarp}, the gravitino spectrum \cite{leepapa}, cosmological
de-Sitter or scaling solutions \cite{cosmologynew}\footnote{See
Ref. \cite{cosmologyold} for
 old cosmological solutions without the presence of  branes.}, regularisation of the
conical singularities \cite{thickb,regular}, cosmology on a
regularised brane\cite{regbranecos1,regbranecos2}, modulus
stabilisation \cite{modulustab}, the Casimir effect \cite{mina},
the effective 4D theory using the gradient expansion
\cite{gradient}, exact wave solutions \cite{waves}, etc. In the
literature, however, the branes are regarded as breaking SUSY
explicitly at the scale of brane tensions.

In this paper, we consider the supersymmetrisation  of the brane
tension action in a way compatible with the bulk SUSY in 6D
Salam-Sezgin supergravity. We find that a brane-localised
Fayet-Ilioupolos (FI) 
term\footnote{An arbitrary brane-localised FI term was considered to see the effect
on the quantization condition in Refs.~\cite{otherwarped,finetune1}.
In 6D global SUSY, the effect of the FI term on the localisation and 
the Kaluza-Klein(KK) mass spectrum of bulk fields was discussed
in Ref.~\cite{lnz}.} 
proportional to each brane tension must
be introduced to cancel the SUSY variation of the brane tension
term. With a nonzero FI term, we should also add in the action the
brane-localised bilinear fermion terms that couple to the $U(1)_R$
field strength. Furthermore, we should modify the SUSY
transformation of the $U(1)_R$ gaugino with a singular term. The
$Z_2$ orbifold boundary conditions on the branes are also required
to project out half of the bulk SUSY.
In order to get the right Bianchi identities with the modified gauge field strengths,
we also need to add a localised correction to the Chern-Simons term in the field strength for
the Kalb-Ramond field appearing in the action and the SUSY transformation.

Consequently, solving the modified equations of motion  with
singular FI terms, we find that the axisymmetric warped solution
of the non-SUSY brane action is maintained, because the localised
FI term is cancelled by a singular piece of the $U(1)_R$ field
strength. However, the Wilson line phase of the gauge potential is
now fixed to be nonzero at the brane position due to the extra
singular term in the gauge field equation. From the SUSY
variations of the spinors, we show that the only supersymmetric
solution with branes is the unwarped "football"-shaped
compactification. Moreover, we find that the FI terms change
the flux quantization condition such that the brane tensions 
are not quantized any more for the same monopole
number as in the Salam-Sezgin vacuum\cite{SS}.
Furthermore, the FI terms affect the number of zero modes of gravitino and we expect that
the same is true for any $U(1)_R$ charged bulk field.

By analysing the equation for the 4D component gravitino, we show
that even after the $Z_2$ projection around the branes, there are
generically multiple normalizable zero modes of the gravitino. In
particular, for the "football" solutions, there are multiple
chiral zero modes only from the left-handed gravitino: the one
with zero winding number and pairs of chiral zero modes with
nonzero winding numbers $(m,-m)$. The mass terms for them would be
forbidden unless the two $U(1)$ gauge symmetries in the system,
the $U(1)_Q$ isometry of the axisymmetric extra dimensions and the
$U(1)_R$ symmetry, are broken. In this "football" case, we propose
that it is possible to have only one chiral zero mode of the 4D
gravitino left (with zero winding number), if a linear combination
of the $U(1)$ symmetries remains unbroken at low energies. The
survival of only one chiral gravitino would be what one should
expect from 4D unbroken ${\cal N}=1$ supergravity.

The paper is organized as follows. First we present the bulk
action of the 6D Salam-Sezgin supergravity to fix the notations.
Then we consider the supersymmetrisation of the brane tension
action and derive the required supersymmetric brane-bulk
couplings. We go on to discuss the modified solutions with the
localised FI terms, identify the supersymmetric football-shaped
solution and study the effect on the zero modes of gravitino.
Finally, the conclusions are drawn.

\section{Six-dimensional Salam-Sezgin supergravity}

The six-dimensional Salam-Sezgin supergravity \cite{SS} consists
of gravity coupled to a dilaton field $\f$, a $U(1)_R$ gauge field
$A_M$ and a Kalb-Ramond field $B_{MN}$, along with the necessary
SUSY fermionic fields, the gravitino $\psi_M$, the dilatino $\chi$
and the gaugino $\la$  where all spinors are 6D Weyl.
The $U(1)_R$ gauge
field corresponds to the gauging of the $R$-symmetry of
six-dimensional supergravity. The complete bulk Langrangian up to
four fermion terms is given by
 \bea e^{-1}_6{\cal L}_{\rm
bulk}&=&R-\frac{1}{4}(\partial_M\phi)^2-\frac{1}{12}
e^{\phi}G_{MNP}G^{MNP}
-\frac{1}{4}e^{\frac{1}{2}\phi}F_{MN}F^{MN}-8g^2 e^{-\frac{1}{2}\phi} \nonumber \\
&&+{\bar\psi}_M\Gamma^{MNP}{\cal D}_N\psi_P+
{\bar\chi}\Gamma^M{\cal D}_M\chi
+{\bar\lambda}\Gamma^M{\cal D}_M\lambda\nonumber\\
&&+\frac{1}{4}(\partial_M\phi)({\bar\psi}_N\Gamma^M\Gamma^N\chi +
\bar{\chi}\Gamma^N\Gamma^M\psi_N)\nonumber \\
&&+\frac{1}{24} e^{\frac{1}{2}\phi}G_{MNP}(\bar{\psi}^R\Gamma_{[R}\Gamma^{MNP}\Gamma_{S]}\psi^S
+\bar{\psi}_R\Gamma^{MNP}\Gamma^R\chi \nonumber \\
&&\quad-\bar{\chi}\Gamma^R\Gamma^{MNP}\psi_R
-{\bar\chi}\Gamma^{MNP}\chi+\bar{\lambda}\Gamma^{MNP}\lambda) \nonumber \\
&&-\frac{1}{4\sqrt{2}}e^{\frac{1}{4}\phi}F_{MN}({\bar\psi}_Q\Gamma^{MN}\Gamma^Q
\lambda+\bar{\lambda}\Gamma^Q\Gamma^{MN}\psi_Q+{\bar\chi}\Gamma^{MN}\lambda-\bar{\lambda}\Gamma^{MN}\chi) \nonumber \\
&&+i\sqrt{2}g e^{-\frac{1}{4}\phi}({\bar\psi}_M\Gamma^M\lambda
+\bar{\lambda}\Gamma^M\psi_M-{\bar\chi}\lambda+\bar{\lambda}\chi).
\eea The field strengths of the gauge and the Kalb-Ramond fields
are defined as \bea
F_{MN}&=& \partial_M A_N - \partial_N A_N, \\
G_{MNP}&=& 3\partial_{[M}B_{NP]} +\frac{3}{2} F_{[MN}A_{P]},
\label{gmn} \eea and satisfy  the Bianchi identities \bea
\partial_{[Q}F_{MN]}&=&0, \\
\partial_{[Q}G_{MNP]}&=&\frac{3}{4}F_{[MN}F_{QP]}.
\eea For $\delta A_M =\partial_M \Lambda$ under the $U(1)_R$, the
Kalb-Ramond field $B_{MN}$ transforms as \be \delta B_{MN} =
-\frac{1}{2}\Lambda F_{MN}. \ee All the spinors have the same charge
normalized to $+1$ under $U(1)_R$, so the covariant derivative of
the gravitino, for instance, is given by \be {\cal D}_M\psi_N =
(\partial_M+\frac{1}{4}\omega_{MAB}\Gamma^{AB}-igA_M)\psi_N. \ee
The action for this Lagrangian is invariant under the following
local ${\cal N}=2$ SUSY transformations (up to the trilinear
fermion terms): \bea
\delta e^A_M &=& \frac{1}{4}(-\bar{\vep}\Gamma^A\psi_M+\bar{\psi}_M\Gamma^A\vep), \label{susyt1}\\
\delta\phi &=& \frac{1}{2}(\bar{\vep}\chi +\bar{\chi}\vep), \label{susyt2}\\
\delta B_{MN} &=& A_{[M}\delta A_{N]}+\frac{1}{4}e^{-\frac{1}{2}\phi}
(\bar{\vep}\Gamma_M\psi_N-\bar{\psi}_N\Gamma_M\vep-\bar{\vep}\Gamma_N\psi_M+\bar{\psi}_M\Gamma_N\vep
\nonumber\\
&&\quad + \bar{\vep}\Gamma_{MN}\chi-\bar{\chi}\Gamma_{MN}\vep), \label{susyt3}\\
\delta\chi &=& -\frac{1}{4}(\partial_M\phi)\Gamma^M\vep
+\frac{1}{24}e^{\frac{1}{2}\phi}G_{MNP}\Gamma^{MNP}\vep, \label{susyt4} \\
\delta\psi_M &=& {\cal D}_M\vep +\frac{1}{48}e^{\frac{1}{2}\phi}G_{PQR}\Gamma^{PQR}\Gamma_M\vep, \label{susyt5} \\
\delta A_M &=& \frac{1}{2\sqrt{2}} e^{-\frac{1}{4}\phi}(\bar{\vep}\Gamma_M\lambda
- \bar{\lambda}\Gamma_M\vep), \label{susyt6} \\
\delta\lambda &=& \frac{1}{4\sqrt{2}} e^{\frac{1}{4}\phi}F_{MN}\Gamma^{MN}\vep
- i\sqrt{2}g\, e^{-\frac{1}{4}\phi}\vep. \label{susyt7}
\eea

The above spinors are chiral with handednesses
\be \G^7 \psi_M = +
\psi_M, \ \ \  \G^7 \chi = - \chi, \ \ \  \G^7 \la = + \la, \ \ \
\G^7 \varepsilon= + \varepsilon.
\ee
Taking into account that $\G^7=\s^3
\otimes {\bf 1}$ (see Appendix A), the 6D (8-component) spinors
can be decomposed to 6D Weyl (4-component) spinors as
\be \psi_M=
(\tilde{\psi}_M,0)^T, \ \ \  \chi =(0,\tilde{\chi})^T, \ \ \
\lambda= (\tilde{\la},0)^T, \ \ \ \varepsilon = (\tilde{\varepsilon},0)^T.
\ee
For later use, we decompose the 6D Weyl spinor
$\tilde{\psi}$ to $\tilde{\psi}=(\tilde{\psi}_{L},\tilde{\psi}_{ R})^T$,
satisfying $\gamma^5(\tilde{\psi}_{ L},0)^T=+(\tilde{\psi}_{ L},0)^T$ and
$\gamma^5 (0,\tilde{\psi}_{ R})^T=-(0,\tilde{\psi}_{ R})^T$.

\section{Supersymmetrising the brane tension action}

In this section, we will add in the previous action
codimension-two branes with nonzero tension. With this addition,
the total action is no longer invariant under the transformations
\reef{susyt1}-\reef{susyt7}.  We will, thus, modify our action and
SUSY transformations, so that the brane-bulk system is rendered
 supersymmetric. With the modification that we propose, we show that
the bulk action remains supersymmetric while the brane action preserves
${\cal N}=1$ SUSY.

\subsection{Requirements for the supersymmetric brane action}

Let us add to the bulk Lagrangian a term for a brane located at
the position  $y=y_i$, where $y$ is the internal space 2D
coordinate. This  brane Lagrangian will be given by \be {\cal
L}_{\rm brane}=-e_4 T_i \delta^{(2)}(y-y_i), \ee where $T_i$ is
the brane tension and the $2D$ delta function is defined as $\int
d^2y \delta^{(2)}(y-y_i)=1$.

The SUSY transformation of the brane action is non-vanishing as
follows, \be \delta {\cal L}_{\rm brane}=-e_4
\frac{1}{4}T_i\delta^{(2)}(y-y_i) (\bar{\psi}_\mu \Gamma^\mu\vep
+{\rm h.c.}). \ee On the other hand, because the gravitino is
charged under $U(1)_R$, varying the gravitino kinetic term under
\reef{susyt5}, it contains a piece of the gauge field strength as
\bea \delta{\cal L}_{\rm gravitino} &\supset&
e_6\bar{\psi}_M\Gamma^{MNP}{\cal D}_N{\cal D}_P\vep \nonumber
\\
&&\quad=-\frac{i}{2}e_6 g\bar{\psi}_M\Gamma^{MNP}\vep F_{NP}+\cdots.
\eea

We can utilise the above term of the gravitino vatiation to cancel
the brane tension term as following. The $U(1)_R$ field can have
in principle FI localised terms\cite{otherwarped,finetune1} parameterized by constants
$\xi_i$. We can then define a hatted field strength ${\hat
F}_{MN}$  \bea
{\hat F}_{\mu\nu}&=&F_{\mu\nu}, \ \ {\hat F}_{\mu m}=F_{\mu m}, \\
{\hat
F}_{mn}&=&F_{mn}-\epsilon_{mn}\xi_i\,\frac{\delta^{(2)}(y-y_i)}{e_2},
\eea where $\ep_{mn}$ is the 2D volume form, and rewrite the
variation of the gravitino kinetic term as \bea \delta{\cal
L}_{\rm gravitino}&\supset& -\frac{i}{2}e_6
g\bar{\psi}_M\Gamma^{MNP}\vep {\hat F}_{NP}
\nonumber \\
&&\quad +e_4
g\xi_i\delta^{(2)}(y-y_i){\bar\psi}_\mu\Gamma^\mu\gamma^5\vep+\cdots,
\label{gravitinovar} \eea where use is made of
$\Gamma^{mn}\epsilon_{mn}=2\Gamma^{56}=2i\sigma^3\otimes
\gamma^5$, the 6D  chirality condition, $\sigma^3 \otimes {\bf
1}\vep=\vep$, and $\frac{e_6}{e_2}=e_4$. Then, the first term
cancels the variation of the bulk fermion bilinear term, if the
$F_{MN}$ in the fermion bilinear term is replaced with
$\hat{F}_{MN}$. Most importantly, the second term has the right
form to cancel the variation of the brane tension term. The
condition for this to happen is that, \bea
\Big(\gamma_5-\frac{T_i}{4g\xi_i}\Big)\vep (y_i)=0. \eea In other
words, decomposing the SUSY variation spinor as
$\vep=(\tilde{\vep},0)^T$ with
$\tilde{\vep}=(\tilde{\vep}_L,\tilde{\vep}_R)^T$, the following
should be satisfied, \bea
\Big(1-\frac{T_i}{4g\xi_i}\Big)\tilde{\vep}_L(y_i)&=&0, \\
\Big(1+\frac{T_i}{4g\xi_i}\Big)\tilde{\vep}_R(y_i)&=&0.
\eea

Thus, fixing the FI terms with the brane tensions as
$\xi_i=\frac{T_i}{4g}$ or $-\frac{T_i}{4g}$, one needs to impose
that either $\tilde{\vep}_R$ or $\tilde{\vep}_L$  vanish on the
brane. Therefore, only ${\cal N}=1$ SUSY can be preserved on the
brane. For other values of $\xi_i$, both $\tilde{\vep}_L$ and
$\tilde{\vep}_R$ must vanish at the brane, so there would be no
SUSY left. In the bulk action and the SUSY transformations, 
when $F_{MN}$ is replaced by ${\hat F}_{MN}$, 
we also need to modify the field strength $G_{MNP}$ by ${\hat G}_{MNP}$
as
\bea
{\hat G}_{\mu\nu\lambda}&=&G_{\mu\nu\lambda}, \\
{\hat G}_{\mu mn}&=& 3\partial_{[\mu}B_{mn]}+\frac{3}{2}F_{[mn}A_{\mu]}-\xi_iA_\mu\epsilon_{mn}\frac{\delta^{(2)}(y-y_i)}{e_2} \nonumber \\
&=&{\hat G}_{mn\mu}={\hat G}_{n\mu m}.
\eea
On the other hand, keeping the
form of terms $A_M$ to be the
same\footnote{We note, however, that the {\it solutions} for the
gauge field and the Kalb-Ramond field can be changed due to the
singular FI term  compared to the case with no branes, as will be
shown later.} as in the case with no branes, the modified bulk
action is supersymmetric up to four fermion terms.

From now on, we choose $\xi_i=\frac{T_i}{4g}$ for all
branes\footnote{When there are different FI terms on the branes,
there is no SUSY left, which corresponds to an explicit SUSY
breaking by orbifolding.} present in the internal
space, so that there is ${\cal N}=1$ SUSY remaining in the brane action
with a SUSY parameter $\tilde{\vep}_L$ non vanishing on the
branes. This choice is made to agree with the no-brane
Salam-Sezgin vacuum \cite{SS} where a constant $\tilde{\vep}_L$ is
a Killing spinor.

\subsection{Orbifold boundary conditions}

Once an FI term has been chosen to make the brane tension action
invariant under the SUSY transformations, one has in addition to
impose that $\tilde{\vep}_R$ vanishes at the brane position to
preserve ${\cal N}=1$ SUSY on the brane. This can be easily
accomplished if we assume an orbifold $Z_2$ symmetry around the
brane.

If the local complex coordinate around the brane is $z$ (in
locally polar coordinates $z=re^{i \th}$), then the $Z_2$ symmetry
corresponds to \be z \leftrightarrow -z \ \ \ \ \ \ ({\rm or} \ \
\th \leftrightarrow \th + \pi). \ee 
The same $Z_2$ was also introduced in
\cite{salvio} to avoid the possible instability of a negative tension brane. 
We should then assign $Z_2$ parities to all bulk
fields and, of course, the SUSY variation parameters
$\tilde{\vep}_L$ and $\tilde{\vep}_R$. A consistent choice of
parities for the fields and the SUSY variation parameter is \bea
{\rm even}&:&~ \tilde{\psi}_{\alpha L}, \ \tilde{\psi}_{a R}, \
\tilde{\lambda}_L, \
\tilde{\chi}_R, \ \tilde{\varepsilon}_L,\ A_\alpha,\ B_{\alpha\beta}, \ B_{ab}, \ \phi,\\
{\rm odd}&:&~  \tilde{\psi}_{\alpha R},\ \tilde{\psi}_{a L},\
\tilde{\lambda}_R,\
\tilde{\chi}_L, \ \tilde{\varepsilon}_R,\ A_a,\ B_{\alpha a}. \eea where
the gauge field, the Kalb-Ramond field and the gravitino have been
written with locally flat indices, \eg $A_A=e_A^{~M}A_M$, so that
the parity assignments do not depend on the coordinate system. It
is obvious that the above choice of parities forces
$\tilde{\vep}_R$ to vanish on the brane position.

In the case with two branes system, the warped vacua of
 \cite{gibbons} have an axially symmetric internal space. The above
 $Z_2$ symmetry about both branes present, is just a discrete subgroup
  of the axial symmetry. On the other
hand, for the general warped solutions with multiple branes
\cite{leelud}, we require the holomorphic function $V(z)$ in the
metric to satisfy the condition $|V(-z+z_i)|=|V(z-z_i)|$, where
$z_i$ is the $i$-th brane position.

\subsection{The supersymmetric brane-bulk coupling}

As a consequence of introducing the localised FI terms,
we have seen that the brane tension action is made compatible with the
bulk SUSY transformations.
The supersymmetric action of the brane-bulk system up to four fermion terms is
\bea
e^{-1}_6\cal{L}_{\rm SUSY}
&=&R-\frac{1}{4}(\partial_M\phi)^2-\frac{1}{12} e^{\phi}{\hat G}_{MNP}{\hat G}^{MNP}
-\frac{1}{4}e^{\frac{1}{2}\phi}{\hat F}_{MN}{\hat F}^{MN}-8g^2 e^{-\frac{1}{2}\phi} \nonumber \\
&&+{\bar\psi}_M\Gamma^{MNP}{\cal D}_N\psi_P+
{\bar\chi}\Gamma^M{\cal D}_M\chi
+{\bar\lambda}\Gamma^M{\cal D}_M\lambda\nonumber\\
&&+\frac{1}{4}(\partial_M\phi)({\bar\psi}_N\Gamma^M\Gamma^N\chi +
\bar{\chi}\Gamma^N\Gamma^M\psi_N)\nonumber \\
&&+\frac{1}{24} e^{\frac{1}{2}\phi}{\hat G}_{MNP}(\bar{\psi}^R\Gamma_{[R}\Gamma^{MNP}\Gamma_{S]}\psi^S
+\bar{\psi}_R\Gamma^{MNP}\Gamma^R\chi \nonumber \\
&&\quad-\bar{\chi}\Gamma^R\Gamma^{MNP}\psi_R
-{\bar\chi}\Gamma^{MNP}\chi+\bar{\lambda}\Gamma^{MNP}\lambda) \nonumber \\
&&-\frac{1}{4\sqrt{2}}e^{\frac{1}{4}\phi}{\hat F}_{MN}({\bar\psi}_Q\Gamma^{MN}\Gamma^Q
\lambda+\bar{\lambda}\Gamma^Q\Gamma^{MN}\psi_Q+{\bar\chi}\Gamma^{MN}\lambda-\bar{\lambda}\Gamma^{MN}\chi) \nonumber \\
&&+i\sqrt{2}g e^{-\frac{1}{4}\phi}({\bar\psi}_M\Gamma^M\lambda
+\bar{\lambda}\Gamma^M\psi_M-{\bar\chi}\lambda+\bar{\lambda}\chi) \nonumber \\
&&-\frac{e_4}{e_6} \,T_i\delta^{(2)}(y-y_i), \label{mbulkaction}
\eea where the modified gauge field strengths are 
\bea 
{\hat F}_{MN}&=&F_{MN}-\delta^m_M\delta^n_N\epsilon_{mn}\xi_i\frac{\delta^{(2)}(y-y_i)}{e_2},
\label{mfmn} \\
{\hat G}_{MNP}&=&G_{MNP}-3\delta^\mu_{[M} \delta^m_N\delta^n_{P]}A_\mu\epsilon_{mn}\xi_i\frac{\delta^{(2)}(y-y_i)}{e_2},\label{mg}
\eea 
with 
\be 
\xi_i=\frac{T_i}{4g}.\label{fitension}
\ee 
All the fermionic SUSY
transformations are modified as 
\bea 
\delta\chi &=& -\frac{1}{4}(\partial_M\phi)\Gamma^M\vep
+\frac{1}{24}e^{\frac{1}{2}\phi}{\hat G}_{MNP}\Gamma^{MNP}\vep, \label{susyt4a} \\
\delta\psi_M &=& {\cal D}_M\vep +\frac{1}{48}e^{\frac{1}{2}\phi}{\hat G}_{PQR}\Gamma^{PQR}\Gamma_M\vep, \label{susyt5a} \\
\delta\lambda &=&
\frac{1}{4\sqrt{2}} e^{\frac{1}{4}\phi}\hat{F}_{MN}\Gamma^{MN}\vep
- i\sqrt{2}g\, e^{-\frac{1}{4}\phi}\vep,
\label{susyt7a} 
\eea 
but the bosonic SUSY transformations are the same as
eqs.~(\ref{susyt1})-(\ref{susyt3}) and (\ref{susyt6}). 
The important ingredient of
the above modifications is that we have  a brane term linear in
$F_{MN}$, the brane-localised FI term. In other words, there is a
brane coupling to the magnetic flux, which is proportional to the
brane tension. 
Moreover, we get a singular correction to the Chern-Simons term in the field strength for
the KR field. 
We note that the modified field strengths satisfy the Bianchi identities,
$\partial_{[Q}{\hat F}_{MN]}=0$ and $\partial_{[Q}{\hat G}_{MNP]}=\frac{3}{4}{\hat F}_{[MN}{\hat F}_{QP]}$, even with the singular term.

One could be worried by the squared terms of the
two-dimensional delta functions appearing in the kinetic term
$\hat{F}_{MN}\hat{F}^{MN}$. However, SUSY requires these terms to
be present and are a usual ingredient of orbifold supersymmetric
theories \cite{deltasquare,lnz}. 
The delta squared  terms, \ie $\delta^2(0)$, appear naturally
in orbifolds, when bulk and brane fields are coupled
supersymmetrically. One can obtain the same form
$\hat{F}_{MN}\hat{F}^{MN}$ in a  6D off-shell supersymmetric
$U(1)$ theory on $T^2/Z_2$, after the auxiliary field of the bulk
vector multiplet is eliminated\cite{lnz}. It has been known that
the $\delta^2(0)$ term provides counterterms, which are necessary
to maintain supersymmetry in explicit calculations on orbifolds,
like the scattering amplitude and the self-energy correction for a
brane field \cite{deltasquare}. In our case, we have not introduced
brane multiplets other than the tension. The case with brane multiplets
will be studied elsewhere so the usual discussion on the $\delta^2(0)$ term
on orbifolds is expected to hold.

As will be shown in the next section,
when one looks for the solutions of the equations of motion of the
above system, the singular term in the modified gauge field
strength is cancelled by the singular part of the background value
of $F_{MN}$, without affecting the solution of the metric and the dilaton
obtained for the non-SUSY brane action. Only the linear term in $\hat{F}_{MN}$
with arbitrary coefficient has been considered for the non-SUSY brane action\cite{otherwarped,finetune1}. However, in this case, 
even if $F_{MN}$ acquires a singular piece to satisfy the gauge field equation, 
it would lead to a problematic two-dimensional delta
squared term in the Einstein and dilaton equations of motion\cite{finetune1}. 
Moreover, when one looks at the low energy effective
theory, there is a worrisome singular delta squared term
corresponding to the mass term of 4D $U(1)_R$ gauge boson $A_\m$
from ${\hat G}_{\mu m n}{\hat G}^{\mu m n}$. However, by solving the linearized
equation for $B_{MN}$ and inserting the solution for $B_{\mu m}$
into the action, the singular piece of the $B_{\mu m}$ cancels the
contribution of the singular term in ${\hat G}_{\mu m n}$, ending up with the
regular action where the gauge boson gets a finite mass from the
FI terms. Similar cancellations happen in 5D \cite{adam} and 6D
\cite{fll} supergravities coupled to branes.

There are some known anomaly-free models including the
non-abelian gauge fields in 6D gauged supergravity
\cite{RSS,moresugra}. In these cases, an abelian flux can be also
turned on in the direction of the non-abelian gauge fields. For
instance, in the model with $E_7\times E_6\times U(1)_R$ with
hyperino $\bf (912,0)_0$, the $U(1)$ contained in $E_6$ can also
develop a nonzero flux, still maintaining the warped solution that
was obtained for the Salam-Sezgin supergravity \cite{leepapa}. As
a result, $E_6$ is broken down to $SO(10)$ in the bulk and the
adjoint fermions of $E_6$ can survive as two chiral $\bf 16$'s of
$SO(10)$ \cite{RSS}. Even in this more general case, the
supersymmetric brane action obtained for the Salam-Sezgin
supergravity remains the same.

Furthermore, we can always introduce arbitrary localised  FI terms
for any abelian factor\footnote{This does not include U(1)
directions of non-abelian groups, as the one in $E_6$ mentioned
above.} of the bulk gauge group other than $U(1)_R$ in a
supersymmetric way because there is no constraint from the
variation of the gravitino kinetic term unlike
eq.~(\ref{gravitinovar}). We only have to modify the field
strengths appearing in both the bulk action and the fermionic SUSY
transformations like in
eqs.~(\ref{mfmn}), (\ref{mg}) and (\ref{susyt4a})-(\ref{susyt7a}). 
Thus, it is straightforward to see that the
localised FI terms generated in 6D global SUSY case \cite{lnz} are
embedded into a supergravity theory.

\section{Modification of the background solution due to the SUSY-brane action}
\label{solutionSUSY}

In the present section, we will  study the effect of the
brane-localised FI terms to the warped axisymmetric solution that
was obtained for non-SUSY brane action. We will see that the
geometry is not modified by the latter addition, but the gauge
field solution and the quantization condition change.

\subsection{The modified equations of motion}

We will study vacua where the Kalb-Ramond field is consistently
(\ie satisfying its equation of motion) set to zero. Then, the
 Einstein equations  derived from the modified
action (\ref{mbulkaction}) are \bea
R_{MN}=&&2g^2\,e^{-\frac{1}{2}\phi}g_{MN}
+\frac{1}{2}e^{\frac{1}{2}\phi}({\hat F}_{MP}{\hat
F}_N\,^P-\frac{1}{8}g_{MN}{\hat F}^2_{PQ})
\nonumber \\
&&+\frac{1}{4}\partial_M\phi\partial_N\phi+ T^i_{MN}~, \eea where
$T^i_{MN}=-{1 \over 2} {\sqrt{g_4} \over \sqrt{g_6}} T_i
(g^4_{\m\n}\d^\m_M \d^\n_N -g_{MN}) \d^{(2)}(y-y_i)$ is the brane
tension contribution (with $g^4_{\m\n}$ the 4D induced metric).
Furthermore, the dilaton and the gauge field equations read \bea
\square^{(6)}\phi&=&\frac{1}{4}e^{\frac{1}{2}\phi}{\hat F}^2_{PQ}
-8g^2\,e^{-\frac{1}{2}\phi}~,\label{scalareq}
\\
\partial_M(\sqrt{-g}e^{\frac{1}{2}\phi}{\hat F}^{MN})&=&0~.\label{gaugeeq}
\eea

\subsection{The modified warped solution}

Assuming axial symmetry in the internal space, the form of the
general warped solution of \cite{gibbons,otherwarped,otherwarped1} is
maintained, except that the solution for $F_{mn}$ is being replaced
with the hatted one. Thus, the metric, the gauge field and the
dilaton solutions are respectively \bea ds^2&=&W^2(r)
\eta_{\mu\nu}dx^\mu dx^\nu+R^2(r)\bigg(dr^2
+\lambda^2 \Theta^2(r)d\theta^2\bigg), \label{wmetric} \\
{\hat F}_{r\theta}&=&\lambda q  \frac{\Theta R^2}{W^6},  \label{flux}\\
\phi&=&4\ln W ,
\eea with
\bea
R={W \over f_0}, \ \  \ \Theta={r \over W^4},   \\
W^4=\frac{f_1}{f_0}, \ \ f_0=1+\frac{r^2}{r^2_0}, \ \ \
f_1=1+\frac{r^2}{r^2_1},
\eea where $q$ is a  constant denoting
the magnetic flux, and the two radii $r_0$, $r_1$ are given by \be
r^2_0=\frac{1}{2g^2}, \ \ r^2_1=\frac{8}{q^2}. \ee

In the warped solution, the metric has two conical singularities,
one at $r=0$ and the other at $r=\infty$, which is at finite
proper distance from the former one. The  deficit angles
$\delta_i$  of these singularities (supported by brane tensions
$T_i=2\delta_i$) are given by \bea
\frac{\delta_0}{2\pi}&=&1-\lambda, \label{tension1}\\
\frac{\delta_\infty}{2\pi}&=&1-\lambda\frac{r^2_1}{r^2_0}.\label{tension2}
\eea In the  unwarped limit, \ie for $r_0=r_1$, the two brane
tensions must be equal.

Writing the delta function in eq.~(\ref{mfmn}) in polar
coordinates around $r=0$ as
$\d^{(2)}(y-y_i)/e_2=\delta(r)/(2\lambda\pi r)$ and
$\epsilon_{r\theta}=\lambda r$, eq.~(\ref{flux}) becomes \be
F_{r\theta}-{\xi_0 \over 2 \pi}\delta(r)=\lambda q \frac{\Theta
R^2}{W^6}. \ee Then, applying Stokes theorem around the patch
including $r=0$, one obtains that $A_\th(0)=\xi_0/(2\pi)$ and thus
the solution of the only non-zero component of the gauge field is
\be
A_{\theta}=-\frac{4\lambda}{q}\bigg(\frac{1}{f_1}-1\bigg)+\frac{\xi_0}{2\pi
}. \label{gaugep1} \ee 
Likewise, the gauge potential in the patch
surrounding  $r=\infty$ is \be
A_{\theta}=-\frac{4\lambda}{q}\frac{1}{f_1}-\frac{\xi_\infty}{2\pi
}.\label{gaugep2} \ee Hence, after connecting the gauge field
solutions in two patches by a gauge transformation and requiring
that it is single valued under $2\pi$ rotations,  we find the
following  quantization condition should hold \be \frac{4\lambda
g}{q}=n-\frac{g}{2\pi}(\xi_\infty+\xi_0), \ \ n \in {\mathbf Z}.
\label{quantcond} \ee In other words, we find that the FI terms
fix the Wilson line phases of the gauge potential to be
non-vanishing on the branes and can contribute to the quantization
condition for $\xi_0\neq -\xi_\infty$, \ie when $T_0\neq -T_\infty$.
Since the covariant derivative has the same form as in the
case with no branes, the modified background solution for the
gauge potential changes the equations of motions of the other bulk
fields and can affect the number of their zero modes.
Using the flux quantization (\ref{quantcond})
with eqs.~(\ref{tension1}) and (\ref{tension2}), we obtain the
brane tensions are related as \be
\Big(1-\frac{T_0}{4\pi}\Big)\Big(1-\frac{T_\infty}{4\pi}\Big)=\Big[n-
\frac{g}{2\pi}(\xi_\infty+\xi_0)\Big]^2.
\ee
In particular, for the football solution, since $q=4g$ and $\xi_0=\xi_\infty=\frac{\pi}{g}(1-\lambda)$, 
the quantization condition (\ref{quantcond}) is satisfied for $n=1$
and arbitrary $\lambda$.

\section{Supersymmetry of the  background solution}

Calculating  the fermionic SUSY variations \reef{susyt4a},
\reef{susyt5a}, \reef{susyt7a} for the above background solution, we
can find in which cases the background respects or breaks SUSY. In
the general warped background, SUSY is completely broken in the
bulk. This can be seen  just from the SUSY transformation of the
dilatino, \be \delta\chi = -\frac{W'}{W}[\cos\theta
\sigma^1\otimes\gamma^5+\sin\theta \sigma^2\otimes{\bf 1}]\vep,
\ee which is always non-zero. In the special case of zero warping,
\ie when $W'=0$, we need to study the remaining SUSY
transformations.

When there is no brane present, the solution (\ref{wmetric})
becomes a sphere compactification, known as the Salam-Sezgin
vacuum \cite{SS}. The nontrivial SUSY transformations of the
fermions are \bea
\delta\lambda &=& i\sqrt{2} g (\gamma^5-1)\vep, \\
\delta\psi_\theta &=& \bigg[\partial_\theta+i\Big(1-\frac{1}{f_0}\Big)(\gamma^5-1)\bigg]\vep.
\eea
In this case, there exists a constant Killing spinor $\tilde{\vep}_L$, which means that
${\cal N}=1$ SUSY is preserved.

For the "football"-shaped extra dimensions \cite{6dself}, there
are two branes of equal tension, $T_0=T_\infty$, located at the
poles of the sphere. The warp factor is constant, so we have that
$q=4g$ and $n=1$. In this case, the FI
terms make the gauge potential nonzero at the branes and contribute to
the quantization condition. 
In the patch surrounding the brane at
$r=0$, the nontrivial fermionic SUSY transformations are \bea
\delta\lambda &=& i\sqrt{2} g (\gamma^5-1)\vep, \\
\delta\psi_\theta &=& \bigg[\partial_\theta+\frac{i}{2}\bigg\{1+\lambda\Big(1-\frac{2}{f_0}\Big)\bigg\}\gamma^5
+i\lambda\Big(\frac{1}{f_0}-1\Big)-i\frac{g\xi_0}{2\pi}\bigg]\vep \nonumber \\
&=&\bigg[\partial_\theta+\frac{i}{2}\bigg\{1+\lambda\Big(1-\frac{2}{f_0}\Big)\bigg\}(\gamma^5-1)\bigg]\varepsilon,
\eea where use is made of $g\xi_0=\frac{1}{4}T_0=\pi(1-\lambda)$ from
eq.~(\ref{tension1}) in the last line. Then, for a non-zero
left-handed variation parameter $\tilde{\vep}_L$, for which the
gaugino variation is manifestly zero, the remaining nonzero
gravitino variation is \be \delta\tilde{\psi}_{\theta
L}=\partial_\theta\tilde{\vep}_L. \ee So, for any $\lambda$, i.e. any brane tension,
there exists a constant
Killing spinor $\tilde{\vep}_L$ which is $Z_2$-even with respect
to the $r=0$ brane. Thus, we find that the modified spin
connection is cancelled by the nonzero Wilson line
phases at the brane positions, so that ${\cal N}=1$ SUSY is preserved for the
football solution. This is to be compared with the case of
non-SUSY brane action in \cite{leepapa}, where only the case of
odd monopole number $n$ would allow for ${\cal N}=1$ SUSY on the
brane.

\section{The gravitino zero modes}

As we have seen in section \ref{solutionSUSY}  and in particular
in eqs.~(\ref{gaugep1}) and (\ref{gaugep2}),  there are in general
 two possible inequivalent Wilson line phases at the conical
singularities due to the localised FI terms. In this section, we
discuss the effect of these Wilson line phases to the existence of
massless modes of the gravitino. We will also note the differences
from the result obtained in the case for a non-SUSY brane action
\cite{leepapa}.

For comparison with our earlier work \cite{leepapa}, let us move
to a Gaussian normal coordinate system, where the warped solution
is written as \be ds^2=W^2\eta_{\mu\nu}dx^\mu dx^\nu
+d\rho^2+a^2d\theta^2, \ee with $d\rho=R dr, a=\lambda R \Theta$.

After decomposing  the 4D vector part\footnote{We will not be
interested in the extra dimensional vector components of the
gravitino $\psi_m$ which are spin-$\frac{1}{2}$ components.} of
the 6D Weyl gravitino $\psi_\mu=(\tilde{\psi}_\mu,0)^T$ as
$\tilde{\psi}_\mu=(\tilde{\psi}_{\mu L},\tilde{\psi}_{\mu R})^T$
in terms of the 4D Weyl spinors, we make a Fourier expansion of
them as \bea
\tilde{\psi}_{\mu L}&=&\sum_m \tilde{\psi}^{(m)}_{\mu L}(x)\varphi^{(m)}_L(\rho)e^{im\theta}, \\
\tilde{\psi}_{\mu R}&=&\sum_m \tilde{\psi}^{(m)}_{\mu
R}(x)\varphi^{(m)}_R(\rho)e^{im\theta}. \eea
By the redefinition of the 4D gravitino,
there is no mixing of $\tilde{\psi}_\mu$ with the other fermionic
modes \cite{leepapa}. To obtain the massless modes, we set
${\bar\sigma}^\alpha\partial_\alpha \tilde{\psi}^{(m)}_{\mu
L}=\sigma^\beta\partial_\beta \tilde{\psi}^{(m)}_{\mu R}=0$.
Then, the equations of left-handed and right-handed gravitinos are
decoupled \cite{leepapa} and read \bea
\Big[\partial_\rho+\frac{W'}{W}+{1 \over a}(m-\frac{1}{2}\omega-gA_{\theta})\Big]\varphi^{(m)}_R&=&0, \label{zeroeq1} \\
\Big[\partial_\rho+\frac{W'}{W}+{1 \over
a}(-m-\frac{1}{2}\omega+gA_{\theta})\Big]\varphi^{(m)}_L&=&0,
\label{zeroeq2} \eea with $\omega=1-a'$. In the patch surrounding
$r=0$, we can find the explicit solution to the above equations as
\bea
\varphi^{(m)}_L&=& {1 \over W }  ~{\rm exp}\bigg[\int^\rho d\rho' {1 \over a }(m+\frac{1}{2}\omega-gA_{\theta})\bigg] \nonumber \\
&=&{N_m \over W\sqrt{a}}~\Big(\frac{r}{r_0}\Big)^{\frac{s}{2}}~f_0^{\frac{1-t}{2}}, \label{zerow}
\eea
with
\bea
s&=&\frac{1}{\lambda}(1+2m)-\frac{g\xi_0}{\pi \lambda}, \nonumber \\
t&=&\frac{1}{\lambda}(m+\frac{1}{2}-n+\frac{g\xi_\infty}{2\pi})\Big(1-\frac{r^2_0}{r^2_1}\Big)
+\frac{1}{\lambda}\Big[n-\frac{g}{2\pi}(\xi_\infty+\xi_0)\Big]+1,
\eea  
where $N_m$ is the normalization constant.   We note that the
solution for the right-handed gravitino is given by the one for
the left-handed gravitino (\ref{zerow}) with
$(m,n,\xi_0,\xi_\infty)$ being replaced by
$(-m,-n,-\xi_0,-\xi_\infty)$.

From the  normalisation condition  \bea \int d\theta \int d\rho
~Wa~ | \varphi^{(m)}_{L,R}|^2<\infty, \label{nc} \eea we determine
the normalisation constant of the general solution (\ref{zerow})
as \bea N^2_m=\frac{1}{2\pi
r_0}\bigg(\int^\infty_0dx\frac{x^s}{(1+x^2)^t}\bigg)^{-1}\equiv\frac{\Gamma_m}{2\pi
r_0}, \eea with \be
\Gamma_m\equiv\frac{2\Gamma[t]}{\Gamma[(1+s)/2]\Gamma[t-(1+s)/2]}.
\ee Then, in order for  a left-handed zero mode to exist,  the
following  normalisability conditions should be respected, \be
s>-1, \quad s-2t<-1. \ee 
In terms of our original parameters, we
require that \bea - \frac{1}{2}(1+\lambda)+\frac{g\xi_0}{2\pi} < m
< n-\frac{1}{2} \left( 1 - \lambda
\frac{r^2_1}{r^2_0}\right)-\frac{g\xi_\infty}{2\pi}~. \label{norm}
\eea For the right-handed zero mode, the corresponding
normalisability condition reads  \bea
 n+\frac{1}{2} \left( 1 - \lambda \frac{r^2_1}{r^2_0}\right)-\frac{g\xi_\infty}{2\pi} < m < \frac{1}{2}(1+\lambda)+\frac{g\xi_0}{2\pi}.
\label{norm1} \eea Using the relation between the FI term and the
brane  tension (\ref{fitension}), as well as
eqns.~(\ref{tension1}) and (\ref{tension2}), the normalisability
condition becomes for the left-handed mode \be -\lambda < m <
n-1+\lambda \frac{r^2_1}{r^2_0}~,\label{norms1} \ee and for the right-handed mode 
\be n < m < 1.\label{norms2} \ee 
If we
compare the above calculation to the one of the  non-SUSY brane
tensions \cite{leepapa}, we see that in the SUSY brane case, due
to the localised FI terms, there are corrections to the gravitino
wavefunction (\ref{zerow}) and consequently to the normalisability
conditions (\ref{norm}) and (\ref{norm1}). Moreover, it is also
expected that there are modifications to the KK massive modes of the
gravitino\cite{leepapa}.

For the "football"-shaped solutions, we have that $q=4g$ and
$n=1$. For $\lambda=1$, we obtain the well-known Salam-Sezgin
vacuum with one 4D chiral gravitino, the left-handed
 zero mode $\vf_L^{(0)}$. For $\lambda\neq 1$, we see that we will always have normalisable left-handed zero modes $\vf_L^{(m)}$, but
no right-handed ones. The action of the  $Z_2$ parity on the
left-handed modes requires that $m$ is even. Therefore, for $[\lambda]$
even, where $[\lambda]$ is the nearest integer smaller than $\lambda$, 
$([\lambda]-1)$ left-handed zero modes are allowed, and for $[\lambda]$ odd,
$[\lambda]$ left-handed zero modes survive. In all the cases, ${\cal
N}=1$ SUSY is preserved by the background. We note that for $0<\lambda<1$, i.e. 
the positive tension branes, there is only one zero mode coming from the left-handed gravitino.

It would be surprising to find that for $\lambda\geq 3$, the ${\cal N}=1$
unwarped solutions support more than one 4D chiral gravitinos,
because one would expect only one surviving in ${\cal N}=1$ 4D
effective supergravity. The mass terms for these chiral gravitinos
would be forbidden due to the $U(1)$ gauge symmetries: one is the
$U(1)_Q$ isometry of the axisymmetric extra dimensions and the
other is the $U(1)_R$ gauge symmetry\footnote{ Both of them can be
anomaly-free due to the generalised Green-Schwarz mechanism where
the $U(1)$ gauge bosons get masses but the theory is still
invariant due to the axionic coupling to the gauge boson. The
gauge boson mass of the $U(1)_Q$ could be read from a possible
gravitational Chern-Simons term in the three form field strength,
 which arises due to the supersymmetric completion of the Green-Schwarz
term\cite{GSterm}, as in the case of the $U(1)_R$ gauge boson. The
computation of it, is beyond the scope of the present paper.}. The
charge operator ${\hat Q}$ of the $U(1)_Q$ commutes with the 6D
Dirac mass operator \cite{wetterich} and it is given in the 6D
spinor basis by \be {\hat Q}=-i\partial_\theta
+\frac{1}{2}\sigma^3\otimes \gamma^5. \ee

Let us now consider the 4D effective action for the left-handed
zero modes of the gravitino coupled to two $U(1)$ gauge bosons.
The part of the effective low energy Lagrangian that is relevant
in our discussion, is similar with the non-SUSY bulk model
\cite{schaction}, and reads \bea
{\cal L}_{\rm eff}&=&-\frac{1}{4}F^2_{\mu\nu}-\frac{1}{4}F^{'2}_{\mu\nu} \nonumber \\
&&+\sum_{m}{\tilde\psi}^{(m)\dagger}_{\mu L}
{\bar\sigma}^{[\mu}\sigma^\nu{\bar\sigma}^{\lambda]}\Big(\partial_\nu
+\frac{1}{4}\omega_{\nu\alpha\beta}\sigma^{[\alpha}{\bar\sigma}^{\beta]}
-ig_4 R A_\nu -ig'_4 Q A'_\nu\Big) {\tilde\psi}^{(m)}_{\lambda L} \eea
where $A_\mu,A'_\mu$ are the $U(1)_R$ and $U(1)_Q$ gauge bosons
with the 4D effective gauge couplings $g_4$ and $g'_4$, respectively. 
Here, we note that the $R$ and $Q$ charge operators take the values $+1$ and
$m+\frac{1}{2}$ for ${\tilde\psi}^{(m)}_{\mu L}$, respectively.
Then, after changing the basis of the gauge bosons to $A_{1\mu}$
and $A_{2\mu}$ as \bea
A_{1\mu}&=& \frac{1}{\sqrt{4g^2_4+g^{\prime 2}_4}}(g'_4 A_\mu - 2g_4 A'_\mu), \\
A_{2\mu}&=& \frac{1}{\sqrt{4g^2_4+g^{\prime 2}_4}}(2g_4 A_\mu + g'_4 A'_\mu),
\eea
the above action is rewritten as
\bea
{\cal L}_{\rm eff}&=&-\frac{1}{4}F^2_{1\mu\nu}-\frac{1}{4}F^2_{2\mu\nu} \nonumber \\
&&+\sum_{m}{\tilde\psi}^{(m)\dagger}_{\mu L}
{\bar\sigma}^{[\mu}\sigma^\nu{\bar\sigma}^{\lambda]}\Big(\partial_\nu
+\frac{1}{4}\omega_{\nu\alpha\beta}\sigma^{[\alpha}{\bar\sigma}^{\beta]}
-ig_1 Q_1 A_{1\nu} -ig_2 Q_2 A_{2\nu}\Big)
{\tilde\psi}^{(m)}_{\lambda L}, \eea where the new charge
operators are \bea Q_1=R-2Q, \quad Q_2 = \frac{2g^2_4}{g^{\prime 2}_4}
R + Q, \eea and the new gauge couplings are \be
g_1=\frac{g_4g'_4}{\sqrt{4g^2_4+g^{\prime 2}_4}}, \quad
g_2=\frac{g^{\prime 2}_4}{\sqrt{4g^2_4+g^{\prime 2}_4}}. \ee In this
case, we note that the $Q_1$ charge of the left-handed zero mode
with $m$ winding number is $Q_1=-2m$.

Let us now  suppose that at  low energies, only $Q_1$ survives
while $Q_2$ is broken\footnote{If a linear combination $Q_2$ is
anomalous, it could be broken due to the corresponding FI terms
without breaking SUSY.}. Then, for the "football" solutions, after
the $Z_2$ projection, the remaining left-handed zero modes with
nonzero even and opposite $m$ or $Q_1$ charges can be paired up to
make a 4D Dirac spinor \be
\Psi^{(m)}_{\mu}=({\tilde\psi}^{(m)}_{\mu
L},-i\sigma^2{\tilde\psi}^{(-m)*}_{\mu L})^T, \ee
 so that they get
coupled by their Dirac masses. Therefore, there can be only one chiral
massless mode of the gravitino with $m=0$, \ie the zero mode
uncharged under the $U(1)_1$. The above mechanism for pairing the
left-handed modes, relies on the VEV of a complex scalar
field that breaks the $U(1)_2$, with
appropriate quantum numbers which makes a Yukawa coupling with the
left-handed modes  $Q_2$-invariant. If in addition we write down
localised Majorana mass terms on regularised branes \cite{leepapa}
for the chiral $m=0$ massless mode, we can end up with a non-zero
mass 4D Majorana gravitino. In this case, the remaining ${\cal
N}=1$ SUSY should be also broken by nonzero F-terms on the branes.

For the general warped solution, we find that there are
multiple zero modes of left-handed gravitino with even $m$
while there could also exist zero modes of right-handed gravitino with odd $m$.
In this case, the number of zero modes depends on the warping and the monopole number.

In the presence of the localised FI terms, for a
spin-$\frac{1}{2}$ fermion with the same $U(1)_R$ charge as the
gravitino, a similar analysis can be done like in
Ref.~\cite{salvio}. There is a difference from the gravitino case
only by the warp factor dependence of the wavefunction. The
wavefunction of the zero mode is given by eq.~(\ref{zerow}) with
$W$ being replaced by $W^2$. However, for the spin-$\frac{1}{2}$
fermion, the weighting function in the norm integration (\ref{nc})
is changed to $W^3a$, so the normalization condition is the same
as eqs.~(\ref{norms1}) and (\ref{norms2}) in the gravitino case.
Therefore, a spin-$\frac{1}{2}$ fermion has the same spectrum as
the one of the gravitino. Thus, a pair of the spin-$\frac{1}{2}$ 
zero modes with $(m,-m)$ could be regarded as being eaten by a pair of the
zero modes of the gravitino with $(m,-m)$ to make up a massive 4D
Dirac gravitino. Consequently, each massive 4D Dirac gravitino
should be part of an ${\cal N}=1$ massive spin-$\frac{3}{2}$
supermultiplet.

\section{Conclusions}

In this work, we examined the way to supersymmetrise the
Salam-Sezgin model in the presence of codimension-2 branes
carrying only tension. We have modified the brane action by adding
brane localised FI terms and localised corrections to the Chern-Simons term 
and in addition changed the fermionic SUSY
transformations. The resulting brane action
respects ${\cal N}=1$ SUSY, if the FI terms are chosen
appropriately (related to the brane tension) and requires the
presence of a $Z_2$ symmetry to be realised.

The axisymmetric background solution for the above system is the
same for the metric and dilaton fields as for the non-SUSY brane
action system \cite{gibbons,otherwarped,otherwarped1}. However, the gauge field
solution acquires an additional Wilson line contribution. The last
is important when discussing the SUSY of the background solution.
Therefore, we find that the unwarped solution with "football"-shaped
internal space does not need a quantized brane tension due to the flux quantization condition
and it always respects 4D ${\cal N}=1$ SUSY,
in contrast with the non-SUSY brane action system.

The gravitino zero mode equation of motion was then analysed for
the above-mentioned background. We found the conditions for which
left- and right-handed modes are normalisable. We have focused on
the unwarped "football" background case and remarked that always a
left-handed mode survives with zero winding number $m$. 
For positive brane tensions, i.e. $0<\lambda < 1$, 
there is only one zero mode of gravitino as in the 
Salam-Sezgin vacuum. For negative brane tensions with $\lambda\geq 3$,
there are additional chiral zero modes with non-zero even $m$. It is
conceivable that these extra modes, in some cases, can be paired
to Dirac four-dimensional spinors, leaving only one chiral zero
mode in the massless spectrum.

A natural continuation of the present study is to include
${\cal N}=1$ matter multiplets (chiral and vector) on the branes with
couplings to the bulk fields. This would require a regularisation
of the brane, \eg in the lines of \cite{regular}, since the brane source terms
coupled to the bulk fields other than the brane tension would lead to classical
divergences.
Then, it is expected that SUSY will completely fix the couplings of the brane
with the bulk fields. In this way, we can reconsider the issue
of moduli stabilisation\cite{GSterm,moduli,modulustab}
in the specific gauged supergravity with the supersymmetric branes.
Moreover, if the MSSM fields
are localised on one of the branes, one is expected to draw
important conclusions about the supersymmetry breaking
transmission between the bulk and the branes, or between the two
distant branes in the different geometry than a torus.
A generalization of the above study to multibrane systems without
the axial symmetry\cite{leelud} could also be interesting in that respect.

In addition, a necessary work that is important to be done
is the consistency check of our proposal to eliminate the chiral modes of the
gravitino with non-zero winding number $m$. One should study whether it is
possible in the specific model to have one of the two $U(1)$'s
naturally much heavier than the other, thus leaving one gravitino
with a small mass in the low energy spectum.
Moreover, the decoupling of the chiral modes
with non-zero $m$ relies on the nonzero VEV of a scalar field which has
a right quantum number $Q_2$ for the Yukawa coupling.
We plan to investigate the above questions in the near future.

\section*{Acknowledgments}

H.M.L. would like to thank Adam Falkowski for his contribution 
and discussion on the Fayet-Iliopoulos term for a supersymmetric brane.
H.M.L. is supported by the DOE Contracts DOE-ER-40682-143 and
DEAC02-6CH03000. A.P. is supported by a Marie Curie Intra-European
Fellowship EIF-039189.

\def\theequation{A.\arabic{equation}}
\setcounter{equation}{0}
\vskip0.8cm
\noindent
{\Large \bf Appendix A: Notations and conventions}
\vskip0.4cm
\noindent

We use the metric signature $(-,+,+,+,+,+)$ for the 6D metric. The
index conventions are the following: (1) for the Einstein indices
we use  $M,N,\cdots=0,\cdots,5,6$ for the 6D indices,
$\mu,\nu,\cdots,=0,\cdots,3$ for the 4D indices and
$m,n,\cdots=5,6$ for the internal 2D indices,  (2) for the Lorentz
indices we use $A,B,\cdots=0,\cdots,5,6$ for the 6D indices,
$\alpha,\beta,\cdots=0,\cdots,3$  for the 4D indices and
$a,b,\cdots=5,6$  for the internal 2D indices.

We take the gamma matrices in the locally flat
coordinates\cite{SS}, satisfying
$\{\Gamma_A,\Gamma_B\}=2\eta_{AB}$, to be \bea
\Gamma_\alpha&=&\sigma^1\otimes\gamma_\alpha, \ \
\Gamma_5=\sigma^1\otimes\gamma_5, \ \ \Gamma_6=\sigma^2\otimes{\bf
1}, \eea where $\g$'s are the 4D gamma matrices with
$\gamma^2_5=1$ and $\sigma$'s are the  Pauli matrices with
$[\s^i,\s^j]=2i \ep_{ijk} \s^k$, with $i,j,k=1,2,3$, \be
\s^1 = \left(\begin{array}{ll}0 & 1 \\
1 & 0 \end{array}\right), \ \ \ \s^2 = \left(\begin{array}{lr}0 & -i \\
i & 0 \end{array}\right), \ \ \ \s^3 = \left(\begin{array}{lr}1 & 0 \\
0 & -1 \end{array}\right).
\ee
The curved gamma matrices on the other hand are given in terms of the ones in the locally
flat coordinates as $\G^M= e^{~M}_A \G^A$ where $e^{~M}_A$ is the 6D vielbein. In addition, the  6D chirality operator is given by
\be
\Gamma_7=\Gamma_0\Gamma_1\cdots\Gamma_6=\sigma^3\otimes{\bf 1}.
\ee
The convention for 4D gamma matrices is that
\be
\g^\a = \left(\begin{array}{ll}0 & \s^\a \\
\bar{\s}^\a & 0 \end{array}\right), \ \  \g^5 = \left(\begin{array}{lr} {\bf 1} & 0 \\
0 & -{\bf 1} \end{array}\right), \ee with $\s^\a=({\bf 1}, \s^i)$
and $\bar{\s}^\a=(-{\bf 1}, \s^i)$. The chirality projection
operators are defined as $P_L=(1+\gamma^5)/2$ and
$P_R=(1-\gamma^5)/2$.

Finally, some useful quantities which we use in the text are the
following \be \G^{\a 5}={\bf 1} \otimes \g^\a \g^5, \ \ \ \G^{\a
6}= i \s^3 \otimes \g^\a, \ \ \ \G^{56}=i \s^3 \otimes \g^5. \ee


\begin{thebibliography}{99}


\bibitem{branes}
  N.~Arkani-Hamed, S.~Dimopoulos and G.~R.~Dvali,
  Phys.\ Lett.\  B {\bf 429} (1998) 263
  [arXiv:hep-ph/9803315];
  I.~Antoniadis, N.~Arkani-Hamed, S.~Dimopoulos and G.~R.~Dvali,
  Phys.\ Lett.\  B {\bf 436} (1998) 257
  [arXiv:hep-ph/9804398];
  L.~Randall and R.~Sundrum,
  Phys.\ Rev.\ Lett.\  {\bf 83} (1999) 3370
  [arXiv:hep-ph/9905221].

\bibitem{RS}
  L.~Randall and R.~Sundrum,
  Nucl.\ Phys.\  B {\bf 557} (1999) 79
  [arXiv:hep-th/9810155].

\bibitem{LS}
  M.~A.~Luty and R.~Sundrum,
  Phys.\ Rev.\  D {\bf 62} (2000) 035008
  [arXiv:hep-th/9910202].




\bibitem{anomalymed}
  G.~F.~Giudice, M.~A.~Luty, H.~Murayama and R.~Rattazzi,
  JHEP {\bf 9812} (1998) 027
  [arXiv:hep-ph/9810442].


\bibitem{highseq}
  A.~Anisimov, M.~Dine, M.~Graesser and S.~D.~Thomas,
  Phys.\ Rev.\  D {\bf 65} (2002) 105011
  [arXiv:hep-th/0111235];
  A.~Anisimov, M.~Dine, M.~Graesser and S.~D.~Thomas,
  JHEP {\bf 0203} (2002) 036
  [arXiv:hep-th/0201256].


\bibitem{fll}
  A.~Falkowski, H.~M.~Lee and C.~Ludeling,
  JHEP {\bf 0510} (2005) 090
  [arXiv:hep-th/0504091].



\bibitem{symseq}
  M.~Schmaltz and R.~Sundrum,
  JHEP {\bf 0611} (2006) 011
  [arXiv:hep-th/0608051];
  S.~Kachru, L.~McAllister and R.~Sundrum,
  arXiv:hep-th/0703105.





\bibitem{5dwarp}
  A.~Falkowski, Z.~Lalak and S.~Pokorski,
  Phys.\ Lett.\  B {\bf 491} (2000) 172
  [arXiv:hep-th/0004093];
  J.~Bagger, D.~Nemeschansky and R.~J.~Zhang,
  JHEP {\bf 0108} (2001) 057
  [arXiv:hep-th/0012163];
  E.~Bergshoeff, R.~Kallosh and A.~Van Proeyen,
  JHEP {\bf 0010} (2000) 033
  [arXiv:hep-th/0007044];
  J.~Bagger and D.~V.~Belyaev,
  Phys.\ Rev.\  D {\bf 67} (2003) 025004
  [arXiv:hep-th/0206024].



\bibitem{SS}
H.~Nishino and E.~Sezgin,
  Phys.\ Lett.\  B {\bf 144} (1984) 187;
  A.~Salam and E.~Sezgin,
  Phys.\ Lett.\ B {\bf 147} (1984) 47.





\bibitem{gibbons}
G.~W.~Gibbons, R.~Guven and C.~N.~Pope,
  Phys.\ Lett.\ B {\bf 595} (2004) 498
  [arXiv:hep-th/0307238].

\bibitem{otherwarped}
Y.~Aghababaie {\it et al.},
  JHEP {\bf 0309} (2003) 037
  [arXiv:hep-th/0308064].

\bibitem{otherwarped1}
  C.~P.~Burgess, F.~Quevedo, G.~Tasinato and I.~Zavala,
  JHEP {\bf 0411} (2004) 069
  [arXiv:hep-th/0408109].


\bibitem{leelud}
H.~M.~Lee and C.~Ludeling,
  JHEP {\bf 0601} (2006) 062
  [arXiv:hep-th/0510026].


\bibitem{gian}
  S.~L.~Parameswaran, G.~Tasinato and I.~Zavala,
  Nucl.\ Phys.\  B {\bf 737} (2006) 49
  [arXiv:hep-th/0509061].





\bibitem{6dself}
 S.~M.~Carroll and M.~M.~Guica,
arXiv:hep-th/0302067;
I.~Navarro,
JCAP {\bf 0309} (2003) 004 [arXiv:hep-th/0302129];
  Y.~Aghababaie, C.~P.~Burgess, S.~L.~Parameswaran and F.~Quevedo,
  Nucl.\ Phys.\ B {\bf 680} (2004) 389
  [arXiv:hep-th/0304256].



\bibitem{finetune}
I.~Navarro,
Class.\ Quant.\ Grav.\  {\bf 20} (2003) 3603
[arXiv:hep-th/0305014];
H.~P.~Nilles, A.~Papazoglou and G.~Tasinato,
Nucl.\ Phys.\ B {\bf 677} (2004) 405
[arXiv:hep-th/0309042];
J.~Garriga and M.~Porrati,
JHEP {\bf 0408} (2004) 028
[arXiv:hep-th/0406158].


\bibitem{finetune1}
H.~M.~Lee,
Phys.\ Lett.\ B {\bf 587} (2004) 117
[arXiv:hep-th/0309050].



\bibitem{koy}
  K.~Koyama,
  arXiv:0706.1557 [astro-ph].




\bibitem{japs}
  S.~Mukohyama, Y.~Sendouda, H.~Yoshiguchi and S.~Kinoshita,
  JCAP {\bf 0507} (2005) 013
  [arXiv:hep-th/0506050];
H.~Yoshiguchi, S.~Mukohyama, Y.~Sendouda and S.~Kinoshita,
  JCAP {\bf 0603} (2006) 018
  [arXiv:hep-th/0512212].





\bibitem{pert}
   H.~M.~Lee and A.~Papazoglou,
  Nucl.\ Phys.\ B {\bf 747} (2006) 294
  [arXiv:hep-th/0602208],
  Erratum-ibid. B {\bf 765} (2007) 200;
S.~L.~Parameswaran, S.~Randjbar-Daemi and A.~Salvio,
  Nucl.\ Phys.\  B {\bf 767} (2007) 54
  [arXiv:hep-th/0608074];
C.~P.~Burgess, C.~de Rham, D.~Hoover, D.~Mason and A.~J.~Tolley,
  JCAP {\bf 0702} (2007) 009
  [arXiv:hep-th/0610078].


\bibitem{salvio}
  S.~L.~Parameswaran, S.~Randjbar-Daemi and A.~Salvio,
  arXiv:0706.1893 [hep-th].




\bibitem{stabwarp}
  T.~Kobayashi and M.~Minamitsuji,
  Phys.\ Rev.\  D {\bf 75} (2007) 104013
  [arXiv:hep-th/0703029].




\bibitem{leepapa}
  H.~M.~Lee and A.~Papazoglou,
  arXiv:0705.1157 [hep-th].





\bibitem{cosmologynew}
A.~J.~Tolley, C.~P.~Burgess, D.~Hoover and Y.~Aghababaie,
  JHEP {\bf 0603} (2006) 091
  [arXiv:hep-th/0512218];
  A.~J.~Tolley, C.~P.~Burgess, C.~de Rham and D.~Hoover,
  New J.\ Phys.\  {\bf 8} (2006) 324
  [arXiv:hep-th/0608083];
B.~Himmetoglu and M.~Peloso,
  Nucl.\ Phys.\  B {\bf 773} (2007) 84
  [arXiv:hep-th/0612140];
T.~Kobayashi and M.~Minamitsuji,
  arXiv:0705.3500 [hep-th];
E.~J.~Copeland and O.~Seto,
  JHEP {\bf 0708} (2007) 001
  [arXiv:0705.4169 [hep-th]].










\bibitem{cosmologyold}
  K.~i.~Maeda and H.~Nishino,
  Phys.\ Lett.\  B {\bf 154} (1985) 358;
  Phys.\ Lett.\  B {\bf 158} (1985) 381.


\bibitem{thickb}
B.~Carter, R.~A.~Battye and J.~P.~Uzan,
  Commun.\ Math.\ Phys.\  {\bf 235} (2003) 289
  [arXiv:hep-th/0204042];
M.~Kolanovic, M.~Porrati and J.~W.~Rombouts,
  Phys.\ Rev.\ D {\bf 68} (2003) 064018
  [arXiv:hep-th/0304148];
S.~Kanno and J.~Soda,
  JCAP {\bf 0407} (2004) 002
  [arXiv:hep-th/0404207];
I.~Navarro and J.~Santiago,
  JHEP {\bf 0502} (2005) 007
  [arXiv:hep-th/0411250];
  C.~de Rham,
  arXiv:0707.0884 [hep-th].


\bibitem{regular}
   M.~Peloso, L.~Sorbo and G.~Tasinato,
   Phys.\ Rev.\  D {\bf 73} (2006) 104025
   [arXiv:hep-th/0603026];
   E.~Papantonopoulos, A.~Papazoglou and V.~Zamarias,
   JHEP {\bf 0703} (2007) 002
   [arXiv:hep-th/0611311];
N.~Kaloper and D.~Kiley,
  JHEP {\bf 0705} (2007) 045
  [arXiv:hep-th/0703190].



\bibitem{regbranecos1}
J.~Vinet and J.~M.~Cline,
  Phys.\ Rev.\ D {\bf 70} (2004) 083514
  [arXiv:hep-th/0406141];
J.~Vinet and J.~M.~Cline,
  Phys.\ Rev.\ D {\bf 71} (2005) 064011
  [arXiv:hep-th/0501098].


\bibitem{regbranecos2}
  E.~Papantonopoulos, A.~Papazoglou and V.~Zamarias,
  arXiv:0707.1396 [hep-th];
  M.~Minamitsuji and D.~Langlois,
  arXiv:0707.1426 [hep-th].

\bibitem{modulustab}
C.~P.~Burgess, D.~Hoover and G.~Tasinato,
  JHEP {\bf 0709} (2007) 124
  [arXiv:0705.3212 [hep-th]].



\bibitem{mina}
  E.~Elizalde, M.~Minamitsuji and W.~Naylor,
  Phys.\ Rev.\  D {\bf 75} (2007) 064032
  [arXiv:hep-th/0702098];
  M.~Minamitsuji,
  arXiv:0704.3623 [gr-qc].






\bibitem{gradient}
S.~Fujii, T.~Kobayashi and T.~Shiromizu,
  arXiv:0708.2534 [hep-th];
 F.~Arroja, T.~Kobayashi, K.~Koyama and T.~Shiromizu,
 arXiv:0710.2539 [hep-th].



\bibitem{waves}
  A.~J.~Tolley, C.~P.~Burgess, C.~de Rham and D.~Hoover,
  arXiv:0710.3769 [hep-th].



\bibitem{lnz}
  H.~M.~Lee, H.~P.~Nilles and M.~Zucker,
  Nucl.\ Phys.\  B {\bf 680} (2004) 177
  [arXiv:hep-th/0309195].


\bibitem{deltasquare}
  E.~A.~Mirabelli and M.~E.~Peskin,
  Phys.\ Rev.\  D {\bf 58} (1998) 065002
  [arXiv:hep-th/9712214];
  H.~M.~Lee,
  JHEP {\bf 0506} (2005) 044
  [arXiv:hep-th/0502093].

\bibitem{adam}
 A.~Falkowski,
  JHEP {\bf 0505} (2005) 073
  [arXiv:hep-th/0502072].




\bibitem{RSS}
  S.~Randjbar-Daemi, A.~Salam, E.~Sezgin and J.~A.~Strathdee,
  Phys.\ Lett.\ B {\bf 151} (1985) 351.






\bibitem{moresugra}
  S.~D.~Avramis, A.~Kehagias and S.~Randjbar-Daemi,
  JHEP {\bf 0505} (2005) 057
  [arXiv:hep-th/0504033];
  S.~D.~Avramis and A.~Kehagias,
  JHEP {\bf 0510} (2005) 052
  [arXiv:hep-th/0508172];
  R.~Suzuki and Y.~Tachikawa,
  J.\ Math.\ Phys.\  {\bf 47} (2006) 062302
  [arXiv:hep-th/0512019].









\bibitem{GSterm}
  Y.~Aghababaie, C.~P.~Burgess, S.~L.~Parameswaran and F.~Quevedo,
  JHEP {\bf 0303} (2003) 032
  [arXiv:hep-th/0212091].


\bibitem{wetterich}
  C.~Wetterich,
  Nucl.\ Phys.\  B {\bf 253} (1985) 366;
 J.~M.~Schwindt and C.~Wetterich,
  Phys.\ Lett.\  B {\bf 578} (2004) 409
  [arXiv:hep-th/0309065].


\bibitem{schaction}
  J.~M.~Schwindt and C.~Wetterich,
  Nucl.\ Phys.\  B {\bf 726} (2005) 75
  [arXiv:hep-th/0501049].





\bibitem{moduli}
  A.~P.~Braun, A.~Hebecker and M.~Trapletti,
  JHEP {\bf 0702} (2007) 015
  [arXiv:hep-th/0611102].





\end{thebibliography}
\end{document}